# Nearly Massless Dirac fermions hosted by Sb square net in BaMnSb$_2$

Jinyu Liu[1], Jin Hu[1], Huibo Cao[2], Yanglin Zhu[1], Alyssa Chuang[1], D. Graf[3], D.J. Adams[4], S.M.A. Radmanesh[4], L. Spinu[4], I. Chiorescu[3,5], and Zhiqiang Mao[1*]

[1] Department of Physics and Engineering Physics, Tulane University, New Orleans, LA 70018

[2] Quantum Condensed Matter Division, Oak Ridge National Laboratory, TN 37830

[3] National High Magnetic Field Lab, Tallahassee, FL 32310

[4] Department of Physics and Advanced Materials Research Institute, University of New Orleans, New Orleans, LA 70148

[5] Department of Physics, Florida State University, Tallahassee, FL 32306

## Abstract

Layered compounds AMnBi$_2$ (A=Ca, Sr, Ba, or rare earth element) have been established as Dirac materials. Dirac electrons generated by the two-dimensional (2D) Bi square net in these materials are normally massive due to the presence of a spin-orbital coupling (SOC) induced gap at Dirac nodes. Here we report that the Sb square net in an isostructural compound BaMnSb$_2$ can host nearly massless Dirac fermions. We observed strong Shubnikov-de Haas (SdH) oscillations in this material. From the analyses of the SdH oscillations, we find key signatures of Dirac fermions, including light effective mass (~0.052$m_0$; $m_0$, mass of free electron), high quantum mobility (1280 cm$^2$V$^{-1}$S$^{-1}$) and a $\pi$ Berry phase accumulated along cyclotron orbit. Compared with AMnBi$_2$, BaMnSb$_2$ also exhibits much more significant quasi two-dimensional (2D) electronic structure, with the out-of-plane transport showing nonmetallic conduction below 120K and the ratio of the out-of-plane and in-plane resistivity reaching ~670. Additionally, BaMnSb$_2$ also exhibits a G-type antiferromagnetic order below 283 K. The combination of nearly massless Dirac fermions on quasi-2D planes with a magnetic order makes BaMnSb$_2$ an intriguing platform for seeking novel exotic phenomena of massless Dirac electrons.

*E-mail: zmao@tulane.edu

Three-dimensional topological semimetals, including Dirac semimetals (DSMs)[1–6], Weyl semimetals (WSMs) [7–15] and Dirac nodal-line semimetals [16–21] represent new quantum states of matter and have stimulated intensive studies. These materials possess bulk relativistic quasiparticles with linear energy-momentum dispersion. DSMs feature linear band crossings at discrete Dirac nodes. In WSMs, the Weyl nodes with opposite chirality appear in pairs, and each pair of Weyl nodes can be viewed as evolving from the splitting of Dirac node due to the lifted spin degeneracy arising from either broken spatial inversion symmetry or broken time reversal symmetry (TRS). The linear band dispersions in these materials are topologically protected by crystal symmetry and lead to many distinct physical properties such as large linear magnetoresistance and high bulk carrier mobility [22]. WSMs also show exotic surface "Fermi arc" connecting a pair of Weyl nodes of the opposite chirality [7,8]. These exotic properties of topological semimetals have potential applications in technology.

$AMnBi_2$ (A=alkali earth/rare earth metal) is one of the established Dirac semimetals [23–33]. These materials share common structure characteristics, consisting of alternately stacked $MnBi4$ tetrahedral layers and A-Bi-A sandwich layers [23,26,27,29,31,33]. In an A-Bi-A sandwich layer, Bi atoms form a square net and harbor Dirac fermions, with coincident (e.g. $SrMnBi_2$[34]) or staggered (e.g. $CaMnBi_2$[35]) stacking of A atoms above and below the Bi plane. In the Mn centered edge sharing $MnBi4$ tetrahedral layers, antiferromagnetic (AFM) order usually develops near room temperature [36,37] and such layers are expected to be less conducting [23,27]. Dirac fermions in $AMnBi_2$ have been found to interplay with magnetism, leading to novel exotic properties. This has been demonstrated in $YbMnBi_2$ [15] and $EuMnBi_2$ [31]. Evidence for Weyl state has been observed in $YbMnBi_2$, which has been proposed to be caused by the TRS breaking due to the ferromagnetic (FM) component of a canted AFM state [15]. In $EuMnBi_2$, half-integer bulk quantum Hall Effect (QHE) occurs due to the magnetic order induced two-dimensional (2D) confinement of Dirac fermions [31].

One disadvantage of AMnBi$_2$ as Dirac semimetals is that the strong spin orbit coupling (SOC) due to heavy Bi atoms opens gap at Dirac nodes [23,38], leading to massive Dirac electrons. For instance, in SrMnBi$_2$, the SOC-induced gap at the Dirac node is about 40 meV [23] and the effective mass of Dirac fermions estimated from the analyses of Shubnikov-de Haas (SdH) oscillations is ~ 0.29$m_0$ ($m_0$, the mass of free electron) [23], much heavier than the Dirac fermions in the 3D gapless DSM Cd$_3$As$_2$ where $m^* $ ~ 0.02-0.05$m_0$ [39–42]. Therefore, one possible route to realize massless Dirac fermions in AMnBi$_2$-type material is to replace Bi with other lighter main group elements such as Sb and Sn, whose SOC effect is much weaker. Under this motivation, we previously studied SrMnSb$_2$ [43] and found the 2D Sb layer can indeed harbor much lighter relativistic fermions with $m^*$ ~ 0.14$m_0$. Moreover, this material shows FM properties: the Mn sublattice develops a FM order for 304K < T < 565 K, but a canted AFM state with a FM component for T < 304K. Coupling between ferromagnetism and quantum transport properties of relativistic fermions has also been observed [43]. These interesting results further motivated us to investigate the isostructural compound BaMnSb$_2$, the material studied in this work.

BaMnSb$_2$ crystalizes in a tetragonal structure with the space group of *I4/mmm*[34], similar to the structure of SrMnBi$_2$ but different from the orthorhombic structure of SrMnSb$_2$ [43]. The orthorhombic distortion in SrMnSb$_2$ leads Sb atoms in the Sr-Sb-Sr sandwich layer to form zig-zag chains. However, in BaMnSb$_2$, as Ba has larger ionic radius than Sr, the stronger interaction between Ba atoms and Sb atoms suppresses orthorhombic distortion[38] and lead to a Sb square net lattice (Fig. **1**a), which is an analogue of the Bi square net in SrMnBi$_2$. In addition, the Ba layers are coincidently stacked along the Sb layer in BaMnSb$_2$, which is also distinct from the staggeredly arranged Sr atoms in SrMnSb$_2$[38]. First principle calculations have predicted that BaMnSb$_2$ exhibit Dirac fermion behavior and its SOC induced gap near the Dirac node is as small as ~20 meV, about half of the gap in SrMnBi$_2$[38].

In this paper, we will show BaMnSb$_2$ indeed is a Dirac material with the Dirac node being very close to the Fermi level. Through the analyses of SdH oscillations, we find this material hosts nearly massless Dirac fermions ($m^* \sim 0.052 m_0$). As compared with AMnBi$_2$ and SrMnSb$_2$, BaMnSb$_2$ possesses the smallest Fermi surface (FS) with the most significant 2D character. Additionally, BaMnSb$_2$ also exhibits a G-type AFM order below 283 K. These findings suggest that BaMnSb$_2$ is a promising candidate for seeking novel exotic phenomena of massless Dirac fermions.

**Results and discussions**

The BaMnSb$_2$ single crystals used in this study were synthesized using self-flux method (see Method). The composition measured by an energy dispersive X-ray spectrometer (EDS) can be expressed as Ba$_{1-y}$Mn$_{1-z}$Sb$_2$, with $y < 0.05$ and $0.05 < z < 0.12$. The nonstoichiometry of Sr and Mn was also found in SrMnSb$_2$ where the actual composition measured by EDS can be described by Sr$_{1-y}$Mn$_{1-z}$Sb$_2$ ($y, z < 0.1$)[43]. The neutron diffraction experiment on a piece of single crystal with the measured composition of Ba$_{0.96}$Mn$_{0.94}$Sb$_2$ confirms the tetragonal structure with the space group *I4/mmm* reported by Cordier & Schafer [34]. The lattice parameters and atomic positions extracted from the structural confinement are summarized in Table 1. We note that the lattice constant $c = 23.85$ Å obtained from our structural refinement is smaller than the previously-reported value of 24.34 Å [34], which may be attributed to the deficiencies of Ba and Mn in our sample.

From magnetic susceptibility measurements on BaMnSb$_2$ single crystals, we have observed signatures of antiferromagnetism. As shown in Fig. 1b, at temperatures below 285 K, the susceptibility $\chi$ exhibits a striking decrease for the magnetic field applied along the c-axis (B//c), but a clear upturn for in-plane field (B//ab). As the temperature is lowered below 50K, $\chi$ displays sharp upturns for both field orientations. Similar features have been observed in AMnBi$_2$ (A=Ca, Sr, and Ba) [23,25,26,32,33,36]. Neutron scattering studies on CaMnBi$_2$ and SrMnBi$_2$ have demonstrated that those magnetic anomalies probed in susceptibility originate from an AFM order formed on the Mn-sublattice [36]. Given the similar behavior in susceptibility between BaMnSb$_2$ and AMnBi$_2$, we can reasonably attribute the magnetic transition at 285 K seen in BaMnSb$_2$ to an AFM transition. We note BaMnSb$_2$ exhibits distinct magnetic anisotropy from AMnBi$_2$. As seen in Fig. 1b, $\chi$ of B//c ($\chi_c$) is much larger than $\chi$ of B//ab ($\chi_{ab}$) at temperatures both above and below T$_N$. However, in AMnBi$_2$, $\chi_{ab}$ is almost equal to $\chi_c$ for $T \geq T_N$, but $> \chi_c$ for $T < T_N$ [23,26,32]. Additionally, the isothermal magnetization $M(H)$ of BaMnSb$_2$ (Fig. 1c) also looks different from that of AMnBi$_2$. We have observed weak FM polarization behavior (Fig. 1c), in contrast with the nearly linear field dependence of $M$ for BaMnBi$_2$ [32]. These discrepancies imply that BaMnSb$_2$ and AMnBi$_2$ may not share an identical magnetic structure.

Magnetic structures of CaMnBi$_2$, SrMnBi$_2$ and YbMnBi$_2$ have been determined from neutron scattering experiment [36,37]. Although these materials show similar Néel-type AFM coupling within the plane and the moments are oriented along the c-axis, their interlayer coupling is different. CaMnBi$_2$ (space group *P4/nmm*) and YbMnBi$_2$ (*P4/nmm*) features FM interlayer coupling, whereas SrMnBi$_2$ (*I4/mmm*) is characterized by interlayer AFM coupling

[36,37]. The magnetic structure of BaMnSb$_2$ determined from our neutron scattering experiments is similar to that of SrMnBi$_2$, *i.e.* both the interlayer and intralayer couplings between two nearest moments are AFM, which is also called the nearest neighbor G-type AFM order. The Néel temperature probed in the neutron scattering experiment is ~ 283K, as shown in Fig. 1d which shows the Bragg peak at (101) (dominated by the magnetic scattering, see the inset) as well as the temperature dependence of the (101) Bragg peak intensity. The ordered moment of Mn at 4 K estimated from the magnetic structure refinement at 4K is 3.950(85) $\mu_B$/Mn, comparable to the ordered moments probed in CaMnBi$_2$, SrMnBi$_2$, YbMnBi$_2$ and Sr$_{1-y}$Mn$_{1-z}$Sb$_2$ [36, 37, 43]. As indicated above, the isothermal magnetization of BaMnSb$_2$ exhibits weak FM polarization (Fig. 1c). This feature, together with the upturn of $\chi_{ab}$ below $T_N$, sharp upturns of $\chi_{ab}$ and $\chi_c$ below 50K, and the small irreversibility of $\chi_{ab}$ between field cooling (FC) and zero-field-cooling (ZFC) measurements, is reminiscent of a canted AFM state with a FM component. However, moment canting is generally not expected for a tetragonal structure for symmetry considerations. If the weak ferromagnetism turns out to be intrinsic for BaMnSb$_2$, the possible origin may be associated with its actual nonstoichiometric composition Ba$_{1-y}$Mn$_{1-z}$Sb$_2$ as mentioned above. In our previous studies on orthorhombic SrMnSb$_2$ [43], we have demonstrated a FM component arising from a canted AFM state; the saturated FM moment sensitively depends on Sr and Mn deficiencies, ranging from 0.6 $\mu_B$/Mn to 0.005$\mu_B$/Mn. With this in mind, we can speculate that Ba and Mn deficiencies possibly lead to local orthorhombic distortion, thus resulting in local canted AFM states. However, we have to point out that small FM components cannot be resolved in neutron scattering experiments. Hence it is not surprising to see the absence of FM response in our neutron scattering experiment.

We have also characterized the electronic transport properties of BaMnSb$_2$ single crystals. In Fig. 1e we present both in-plane ($\rho_{in}$) and out-of-plane ($\rho_{out}$) resistivity as a function of temperature, from which we found several signatures distinct from that of (Ca/Sr/Ba)MnBi$_2$ [26,30,32,33]. First, BaMnSb$_2$ shows much stronger electronic anisotropy than (Ca/Sr/Ba)MnBi$_2$, which is manifested in its larger $\rho_{out}/\rho_{in}$ ratio. The $\rho_{out}/\rho_{in}$ ratio at 2K ranges from 15 to 100 for (Ca/Sr/Ba)MnBi$_2$ [26,30,32,33], but rises to 670 for BaMnSb$_2$, which is comparable to the value of $\rho_{out}/\rho_{in}$ (~ 609) seen in SrMnSb$_2$ [43]. Such a large electronic anisotropy of BaMnSb$_2$ suggests its electronic structure is quasi-2D like, which is further confirmed in our measurements of angular dependence of SdH oscillation frequency as shown below. Second, unlike (Ca/Sr/Ba)MnBi$_2$ whose $\rho_{out}(T)$ always exhibits a hump due to a crossover from high-temperature incoherent to low-temperature coherent conduction [26,30,32,33], BaMnSb$_2$ displays an opposite behavior in $\rho_{out}(T)$ (Fig. 1e); a crossover from high-temperature metallic conduction to low-temperature localization is observed, which leads to a broad minimum in $\rho_{out}(T)$ around 120K. The temperature dependence of in-plane resistivity $\rho_{in}(T)$ of BaMnSb$_2$ also differs from that of (Ca/Sr/Ba)MnBi$_2$. (Ca/Sr/Ba)MnBi$_2$ features a quadratic temperature dependence for $\rho_{in}$ in low temperature range [23,26,32,33], while $\rho_{in}$ of BaMnSb$_2$ exhibits localization behavior below 80K but crossovers to metallic behavior below 11K. These differences imply the transport mechanism in BaMnSb$_2$ is somewhat different from that in (Ca/Sr/Ba)MnBi$_2$. We note that in the temperature region where the localization behavior occurs, both $\rho_{out}(T)$ and $\rho_{in}(T)$ follow a logarithmic temperature dependence, as denoted by the dashed lines in Fig. 1f which presents $\rho_{out}(T)$ and $\rho_{in}(T)$ on the log$T$ scale. This observation is reminiscent of Kondo effect. Given that we have an AFM lattice formed from local moments of Mn ions, the presence of Kondo effect is possible in principle. But, this naturally leads to a question why such an effect occurs only to BaMnSb$_2$, but not to

(Ca/Sr/Ba)MnBi$_2$ and SrMnSb$_2$ with similar AFM lattices. Clear understanding of this issue requires further studies, but one possible interpretation is that the Kondo effect depends on the dimensionality of electronic structure, and may be enhanced in BaMnSb$_2$ due to its highly 2D electronic structure. Moreover, we would like to point out the localization behavior seen in BaMnSb$_2$ cannot be attributed to disorder induced localization since in Sr$_{1-y}$Mn$_{1-z}$Sb$_2$ with the level of disorders being comparable or higher than that of BaMnSb$_2$, no localization behavior is observed [43].

Like (Ca/Sr/Ba)MnBi$_2$ and SrMnSb$_2$, BaMnSb$_2$ also exhibits quantum transport properties as revealed by our magnetotransport measurements, In Fig. 2a and 2d, we present the field dependences of in-plane ($\rho_{in}$) and out-of-plane ($\rho_{out}$) resistivity measured at various temperatures for BaMnSb$_2$, respectively. Strong SdH oscillations, which sustain up to above 40K, are observed in both $\rho_{in}(B)$ and $\rho_{out}(B)$. In Fig. 2b and 2e we present the oscillatory components of $\rho_{in}$ and $\rho_{out}$, respectively. From Fast Fourier Transformation (FFT) analyses of $\Delta\rho_{in}$ and $\Delta\rho_{out}$ (see the insets to Fig. 2c and 2f), we find that the SdH oscillations of $\rho_{in}$ consists of a single frequency (~22T), whereas the oscillations of $\rho_{out}$ include two frequencies (*i.e.* $F_\alpha$~25T and $F_\beta$ ~ 35T). Such a difference is likely caused by the non-stoichiometric composition. As mentioned above, the actual composition of our BaMnSb$_2$ crystals involves Ba and Mn non-stoichiometry, which could lead to slight modification for electronic structure in different samples. To verify this speculation, we have measured many samples and find that their oscillation frequencies indeed show variation, ranging from 20T to 35 T. Given that the quantum oscillation frequency is directly linked to the extremal Fermi surface cross-section area $A_F$ by the Onsager relation $F = (\Phi_0/2\pi^2)A_F$, a small oscillation frequency is generally expected for topological semimetals with the Dirac node being near the Fermi level. We note that the quantum oscillation frequency of

22T probed in our BaMnSb$_2$ crystals is the smallest as compared with AMnBi$_2$ and SrMnSb$_2$, implying that if BaMnSb$_2$ turns out to be a Dirac material, its Dirac band crossing points must be very close to the Fermi level.

Evidence for Dirac fermions in BaMnSb$_2$ has been obtained from the further analyses of the SdH oscillations. As shown in Fig. 2c and 2f, the effective cyclotron mass $m^*$ can be extracted from the fit of the temperature dependence of the normalized FFT peak amplitude to the thermal damping factor of Lifshitz-Kosevich (LK) equation [44], i.e. $\Delta\rho/\rho_0 \propto \alpha T\mu/[\bar{B} * \sinh\left(\frac{\alpha T\mu}{\bar{B}}\right)]$, where $\rho_0$ is the zero field resistivity, and $\alpha = (2\pi^2 k_B m_0)/(\hbar e)$. $\mu$ is the ratio of effective mass of cyclotron motion to the free electron mass. $1/\bar{B}$ is the average inverse field for FFT analysis. We did the FFT within the field 3T-31T range for $\rho_{in}$ and the 5T- 31T range for $\rho_{out}$, with $\bar{B}$ being 5.47T and 8.61T respectively. As seen in Fig. 2c and 2f, the best fits yield $m^* = 0.052m_0$ and $0.058m_0$, respectively, for the SdH oscillations of $\rho_{in}$ and $\rho_{out}$. For $\rho_{out}$, the fit was performed for the component with the oscillation frequency of 35T, whose FFT peak can be clearly resolved. The effective mass of $m^* = 0.052m_0$ and $0.058m_0$ seen in BaMnSb$_2$ is much smaller than that of other known AMnBi$_2$ [23,26,32,37] and SrMnSb$_2$ [43], but comparable to that of gapless Dirac semimetal Cd$_3$As$_2$ [39–42]. Detailed comparisons of $m^*$ as well as other parameters derived from SdH oscillations are shown in Table 2.

To further verify if the nearly massless electrons in BaMnSb$_2$ is of topological nature of Dirac fermions, we extracted the Berry phase accumulated along cyclotron orbit from the analyses of SdH oscillations. Berry phase should be zero for a non-relativistic system with parabolic band dispersion, while a finite value up to $\pi$ is expected for Dirac fermions [45,46]. We present the Landau level (LL) fan diagram constructed from the SdH oscillations of $\rho_{in}$ for

BaMnSb$_2$ in Fig. 3a and 3b, where integer LL indices are assigned to the maxima of $\rho_{in}$. Our definition of LL index is based on the customary practice that integer LL indices are assigned to the minima of conductivity [46,47]. In-plane conductivity $\sigma_{xx}$ can be converted from the longitudinal resistivity $\rho_{xx}$ and the transverse (Hall) resistivity $\rho_{xy}$ using $\sigma_{xx} = \rho_{xx}/(\rho_{xx}^2 + \rho_{xy}^2)$. Since our measured $\rho_{xy}$ (Fig. 3d) is about 1/3-1/4 of $\rho_{xx}$ (Fig. 2a) for $B < 9T$, $\sigma_{xx} \approx 1/\rho_{xx}$, which justifies our definition of LL index. As seen in Fig. 3b, the intercept on the LL index axis obtained from the extrapolation of the linear fit in the fan diagram is 0.53, very close to the expected value of 0.5 for a 2D Dirac system with a π Berry phase. The oscillation frequency derived from the fit is 21.8 T, nearly the same as the frequency obtained from the FFT analyses of the SdH oscillations of $\rho_{in}$ (see the inset to Fig. 2c), suggesting that our linear fit in the fan diagram is reliable [46]. The π Berry phase derived from the above fan diagram analyses clearly indicates that the nearly massless electrons probed in the SdH oscillations are Dirac fermions.

Dirac Fermions are usually characterized by high quantum mobility, as seen in Cd$_3$As$_2$ [22]. This is also seen in BaMnSb$_2$. The quantum mobility is directly related with the quantum relaxation time $\tau_q$ by $\mu_q = e\tau_q/m^*$. $\tau_q$ characterizes quantum life time, the time scale over which a quasiparticle stays in a certain eigenstate. $\tau_q$ can be found from the field damping of quantum oscillation amplitude, i.e., $\Delta\rho/\rho_0 \propto \exp(-\alpha T_D \mu/B) * \alpha T\mu/[B * sinh(\alpha T\mu/B)]$. $T_D$ is the Dingle temperature and is linked with $\tau_q$ by $T_D = \hbar/(2\pi k_B \tau_q)$. With $m^*$ being the known parameter, $\tau_q$ at $T = 2K$ can be extracted through the linear fit of $\ln([B * sinh(\alpha T\mu/B)/\alpha T\mu] * \Delta\rho/\rho_0)$ against $1/B$. As shown in Fig 3c, we have obtained $\tau_q = 3.8 \times 10^{-14}$ s, from which the quantum mobility $\mu_q (=e\tau_q/m^*)$ is estimated to be 1280 cm$^2$V$^{-1}$s$^{-1}$, much higher than that of SrMnBi$_2$ (250 cm$^2$V$^{-1}$s$^{-1}$ [23]) or SrMnSb$_2$ (~570 cm$^2$V$^{-1}$s$^{-1}$ [43]) (see Table 2). In general, the transport mobility is one or two orders of magnitude higher than quantum mobility, since the transport mobility is sensitive only to large angle scattering of carriers, while the quantum mobility is sensitive to both small and large angle scatterings. However, this was not observed in BaMnSb$_2$. Using the Hall coefficient $R_H$ data extracted from Hall resistivity data shown in Fig. 3d, the transport mobility $\mu_{tr}(=R_H/\rho_{xx})$ at 1.8 K is estimated to be ~1300 cm$^2$V$^{-1}$S$^{-1}$ for BaMnSb$_2$, much less than that of SrMnSb$_2$ ($\mu_{tr}$ ~12500 cm$^2$V$^{-1}$S$^{-1}$) at low temperature [43]. The low $\mu_{tr}$ in BaMnSb$_2$ may be associated with the transport localization behavior seen in $\rho_{out}$ and $\rho_{in}$.

The Dirac fermion behavior probed in our experiments for BaMnSb$_2$ is in good agreement with the prediction by first principle calculations that BaMnSb$_2$ is a Dirac material at ambient pressure [38]. Next, we will make more detailed comparisons between the predicted electronic band structure and our experimental observations. First, the Dirac bands are predicted to be generated by the Sb square net plane; thus the Fermi surface formed by Dirac bands is expected to be highly 2D, which is supported by our observations. As shown in Fig. 4a and 4d, systematic evolutions of SdH oscillation patterns for $\rho_{in}$ and $\rho_{out}$ are clearly observed as the magnetic field is rotated from the out-of-plane to the in-plane direction (see the insets to Fig. 4b and 4e for the experiment set-up). The oscillation frequency $F(\theta)$ extracted from the FFT for $\rho_{in}$ measurements can be fitted to $F(\theta) = F(\theta=0°)/\cos\theta$ (Fig. 4c), suggesting that the Fermi surface responsible for SdH oscillations in BaMnSb$_2$ is indeed 2D. However, in $\rho_{out}(B,\theta)$ measurements, we observed two frequency branches. As shown in Fig. 4f, the higher frequency branch also follows $F(\theta) = F(0°)/\cos\theta$, while the lower frequency branch shows a weak angular dependence, suggesting the sample used for $\rho_{in}$ measurements has slightly different morphology in its Fermi surface from the sample used for $\rho_{out}$ measurements, which presumably originates from slightly different non-stoichiometric compositions. The 2D Fermi surface also explains the aforementioned large electronic anisotropy manifested in the large $\rho_{out}/\rho_{in}$ ratio (~ 670). Second, the first principle calculations also predicted that the linear Dirac bands crossing occurs near the middle of ΓM, with the crossing point (*i.e.* the Dirac node) being right above the Fermi level, which implies the Fermi pocket hosting Dirac Fermions should be a hole pocket and small. In addition to the hole pocket enclosing the Dirac nodes, small electron pockets with quasi-linear band dispersion are also predicted to exist at X and Y points. The quantum transport properties of Dirac fermion revealed in our experiments provide strong support to these predictions. As shown in Fig. 3d, our measured Hall resistivity exhibits linear field dependence with positive slopes at all temperatures as well as remarkable SdH oscillations below 50K. These features prove that holes are dominant carriers and responsible for the SdH oscillations. From the SdH oscillation frequency of 22T of $\rho_{in}$, the extremal cross-section area of the Fermi surface is estimated to be ~0.2 nm$^{-2}$, about 0.1% of the total area of the first Brillouin zone, indicating an extremely small Fermi surface, the smallest as compared with AMnBi$_2$ and SrMnSb$_2$ (see Table 2). Third, the SOC-induced gap at the Dirac node in BaMnSb$_2$ was predicted to be half of that in SrMnBi$_2$ due to the weaker SOC of the Sb square net as mentioned above, which should result in lighter Dirac Fermions in BaMnSb$_2$. Our observations of small cyclotron Frequency and effective mass of the Dirac fermions in BaMnSb$_2$ are in line with these predicted results. Furthermore, the Dirac cone in BaMnSb$_2$ was predicted to be anisotropic [38], similar to that of SrMnBi$_2$ and CaMnBi$_2$ [23,27,28]. ARPES experiments are called in to verify it.

The signatures of Dirac fermions in BaMnSb$_2$ imply that materials including 2D Sb square net planes can harbor Dirac electrons. We note many such candidate materials indeed exist, *e.g.* ReAgSb$_2$ (Re=rare earth), for which small mass quasi-particles have been found [48–50]. Recent studies on LaAgSb$_2$ have shown its small mass quasi-particles indeed originate from the

Dirac-cone like band structure formed by Sb $5P_{x,y}$ orbitals [51]. Band structure calculations [52] predicted that the Dirac cone in LaAgSb$_2$ can host nearly massless Dirac fermions with $m^* \sim 0.06m_0$ and has a very small Fermi surface with a quantum oscillation frequency of ~20T. However, these predictions were not seen in experiments [48,52]; the smallest $m^*$ measured in experiments is $0.16m_0$ and the least quantum oscillation frequency is 72 T [48]. Surprisingly, the Dirac electron behavior observed in BaMnSb$_2$ is very close to that predicted for LaAgSb$_2$. Note that the electronic structure of BaMnSb$_2$ is much simpler than that of ReAgSb$_2$. BaMnSb$_2$ exhibits only a single frequency quantum oscillations (~22T) and its transport properties can almost be described by a single-band model, whereas LaAgSb$_2$ possesses a much complicated band structure, showing four frequencies in quantum oscillations [48,49].

Given that BaMnSb$_2$ exhibits a G-type AFM order with possible FM components due to Ba and Mn non-stoichiometry as discussed above, a natural question is whether its FM component can be tuned by changing Ba and Mn non-stoichiometry and coupled to quantum transport properties. In our previous studies on Sr$_{1-y}$Mn$_{1-z}$Sb$_2$, we have shown that its saturated FM moment $M_s$ can be tuned from $0.6\mu_B$/Mn to $0.005\mu_B$/Mn by changing $y$ and $z$ [43]. The samples with heavier Sr deficiencies have larger $M_s$ than the samples with heavier Mn deficiencies. The $M_s$ (~$0.04\mu_B$/Mn at 7T, see Fig. 1c) probed in BaMnSb$_2$ seems comparable to that of type B samples of our previously reported Sr$_{1-y}$Mn$_{1-z}$Sb$_2$ where $M_s$ ~0.04-0.06 $\mu_B$/Mn. Although we have examined many samples, all measured samples show comparable $M_s$. Therefore, it is difficult to find samples with a wide range of $M_s$, which would allow us to examine the coupling between Dirac electron behavior and ferromagnetism as we did for Sr$_{1-y}$Mn$_{1-z}$Sb$_2$ [43].

**Conclusion**

In summary, we have demonstrated that in BaMnSb$_2$ the Sb square net layers with coincident stacking of Ba atoms can host nearly massless Dirac fermions due to the weaker SOC effect of Sb, in contrast with massive Dirac fermions hosted by the Bi square net planes in AMnBi$_2$. Compared with AMnBi$_2$, BaMnSb$_2$ displays much more significant 2D-like electronic band structure, with the out-of-plane transport showing non-coherent conduction below 120K and the $\rho_{out}/\rho_{in}$ ratio reaching ~670. Its quantum transport properties can be almost described by a single band model, consistent with its simple electronic band structure predicted by first principle calculations. In addition, BaMnSb$_2$ also exhibits a G-type AFM order below 283K and the Ba and Mn non-stoichiometries might cause a weak FM component. These findings establish BaMnSb$_2$ as a promising platform for seeking novel exotic properties of massless Dirac fermions in low dimensions.

**Methods**

**Single crystal growth and characterization.** Single crystals of BaMnSb$_2$ were synthesized using self-flux method with a stoichiometric ratio of Ba pieces, Mn and Sb powder. The starting materials were mixed in a small crucible, sealed into a quartz tube under Argon atmosphere and heated up to 1050 °C in one day. The temperature was maintained at 1050 °C for two days. After that, it was first cooled down to 1000 °C at a fast rate, 50 °C/h, and then followed by a slowly cooling down to 450 °C at a rate 3 °C/h. Subsequently the furnace was turned off for fast cooling. Plate-like crystals with lateral dimensions of several millimeters (see the inset to Fig. 1b) can easily be obtained from the final product. The compositions of the crystals were measured using an energy dispersive X-ray spectrometer (EDS). The measured composition can be expressed as $Ba_{1-y}Mn_{1-z}Sb_2$, with $y < 0.05$ and $0.05 < z < 0.12$. The structure of the single crystals was characterized by an X-ray diffractometer.

**Magnetization and magnetotransport measurements.** The magnetization data were taken by a 7T SQUID magnetometer (Quantum Design). The magnetotransport properties were measured using standard four and five- probe method for longitudinal and Hall resistivity, respectively, in a Physics Property Measurement System (PPMS, Quantum Design), and the 31 T resistive magnet at National High Magnetic Field Laboratory (NHMFL) in Tallahassee.

**Neutron Scattering.** Single-crystal neutron diffraction was performed at the HB-3A Four-circle Diffractometer equipped with a 2D detector at the High Flux Isotope Reactor(HFIR) at ORNL. Neutron wavelength of 1.546 Å was used from a bent perfect Si-220 monochromator [53]. The Rietveld refinement was performed using FullProf [54].

**Acknowledgements**

The work at Tulane is supported by the U.S. Department of Energy under EPSCoR Grant No. DE-SC0012432 with additional support from the Louisiana Board of Regents (support for a graduate student, materials, travel to NHMFL). The work at NHMFL is supported by National

**Author contributions**

The single crystals used in this study were synthesized and characterized by J.Y.L., Y.L.Z., and A.C. The magnetotransport measurements in PPMS were carried out by J.Y.L., D.J.A, S.M.A.R., L.S. and Z.Q.M. The high field measurements at NHMFL were conducted by J.Y.L., D.G., J.H., S.M.A.R., I.C. and Z.Q.M. J.Y.L. and Z.Q.M. analyzed the data and wrote the manuscript. H.C. performed neutron scattering experiments and data analyses. All authors read and commented on the manuscript. The project was supervised by Z.Q.M.

**Competing financial interests:** The authors declare no competing financial interests.

Figure Captions

**Figure 1. Crystal strucutre, magnetic and transport properties of BaMnSb$_2$. a,** Crystal and magnetic structure of BaMnSb$_2$. **b,** Susceptibility as a function of temperature measured with a 2T magnetic field applied along the *c* axis ($\chi_c$) and along the *ab* plane ($\chi_{ab}$) under zero field cooling (ZFC) and field cooling (FC) histories. Red: $\chi_c$ of FC; dark red: $\chi_c$ of ZFC; purple: $\chi_{ab}$ of FC; blue: $\chi_{ab}$ of ZFC. Inset in **b**: an optical image of a typical BaMnSb$_2$ single crystal. **c,** Isothermal magnetization along the *c* axis (red) and along the *ab* plane (blue). **d,** Temperature dependence of the Bragg peak intensity at (101) indicates the magnetic order below 283 K. Inset in **d**: the Bragg peak (101) scanned at the selected temperatures. **e,** In-plane resistivity ($\rho_{in}$) and out-of-plane resistivity ($\rho_{out}$) as a function of temperature under zero magnetic field. **f,** $\rho_{in}$ and $\rho_{out}$ plotted on logarithmic scale.

**Figure 2. Quantum transport properties of BaMnSb$_2$. a,** The in-plane resistivity, $\rho_{in}$, as a function of field up to 31T at different temperatures. **b,** The oscillatory component of $\rho_{in}$ vs. 1/B at different temperatures. **c,** The temperature dependence of the normalized FFT amplitude. The dashed line curve is the fit to the Lifshitz-Kosevich (LK) formula from 2 to 40K. Inset: FFT spectra of $\Delta\rho_{in}(B)$ at different temperatures (the FFT was done in the field range of 3T-31T). **d,** The out-of-plane resistivity, $\rho_{out}$, as a function of field at different temperatures. **e,** The oscillatory component of $\rho_{out}$ vs. 1/B. **f,** The temperature dependences of the normalized FFT amplitude. The dashed line curve is the fit to the LK formula. Inset: FFT spectra of $\Delta\rho_{out}(B)$ at different temperatures (the FFT was done in the field range of 5T-31T).

**Figure 3. Berry phase, quantum mobility and Hall resistivity of BaMnSb$_2$. a,** The oscillatory

component of in-plane resistivity, $\rho_{in}$, vs. 1/B at 2K. As the longitudinal resistivity ($\rho_{in} = \rho_{xx}$) is much larger than transverse resistivity ($\rho_{xy}$), integer Landau level (LL) indices are assigned to the maximum of resistivity (see text). **b,** LL fan diagram. The blue dashed line represents the linear fit. **c,** Dingle plot for the in-plane quantum oscillations $\Delta\rho_{in}$ at 2K. **d,** Hall resistivitivity as a function of field at various temperatures (T=2, 5, 10, 20, 50, 100, 150, 200, 250 and 300K).

**Figure 4. Angular dependences of SdH oscillations and the oscillation frequencies for BaMnSb$_2$. a,** The oscillatory component of in-plane resistivity, $\Delta\rho_{in}$, vs. 1/B measured under different field orientations. The data has been shifted for clarity. **b,** The FFT spectra of $\Delta\rho_{in}(B)$ at different field orientations. Inset: the diagram of the measurement setup; $\theta$ is defined as the angle between the magnetic field and the out-of-plane direction. **c,** The angle dependence of SdH oscillation frequency determined from the FFT of $\Delta\rho_{in}(B)$. The dashed curve is the fit to $F(\theta) = F(0)/\cos\theta$. **d,** The oscillatory component of out-of-plane resistivity, $\Delta\rho_{out}$, vs. 1/B measured under different field orientations. The data has been shifted for clarity. **e,** The FFT spectra of $\Delta\rho_{out}(B)$ at different field orientations. Inset: the diagram of the measurement setup. **f,** The angular dependence of SdH oscillation frequency determined from the FFT of $\Delta\rho_{out}(B)$. The dashed curves are the fits to $F(\theta) = F(0)/\cos\theta$.

Table 1. Lattice parameters and atomic positions of BaMnSb$_2$ determined from neutron diffraction experiments. The refinement was based on 164 reflections. The R-factor is 0.098 and the $\chi^2$ is 0.33.

| | T = 4K | | | |
|---|---|---|---|---|
| Space group: *I4/mmm*, $a = b = 4.4567(54)$ Å, $c = 23.85(25)$ Å | | | | |
| Magnetic moment, 3.950(85) $\mu_B$/Mn | | | | |
| | x | y | z | Site Multiplicity |
| Ba | 0 | 0 | 0.11325(55) | 4 |
| Mn | 0 | 0.5 | 0.25 | 4 |
| Sb-1 | 0 | 0.5 | 0 | 4 |
| Sb-2 | 0 | 0 | 0.31814(56) | 4 |

Table 2. Comparison of the parameters extracted from the SdH oscillations among (Ca/Sr/Ba)MnBi$_2$ and (Sr/Ba)MnSb$_2$, including the oscillation frequency $F$, effective mass $m^*/m_0$, extremal cross-section area of Fermi surface $A_F$ ($=2\pi^2 F/\Phi_0$), quantum relaxation time $\tau_q$, quantum mobility $\mu_q$ ($=e\tau_q/m^*$), and Berry phase.

| | $F$ (T) | $m^*/m_0$ | $A_F$ (nm$^{-2}$) | $\tau_q$ (s) | $\mu_q$ (cm$^2$V$^{-1}$s$^{-1}$) | Berry's phase | reference |
|---|---|---|---|---|---|---|---|
| CaMnBi$_2$ | 101, 185 | 0.35 | 1.73 | - | - | 0.9$\pi$ | 25, 26 |
| SrMnBi$_2$ | 152 | 0.29 | 1.45 | 3.5×10$^{-14}$ | 250 | 1.2$\pi$ | 23 |
| BaMnBi$_2$ | 33, 83 | 0.105 | - | - | - | 0.4-0.6$\pi$ | 32, 33 |
| SrMnSb$_2$ | 67 | 0.14 | 0.64 | 4.2×10$^{-14}$ | 570 | 1.02$\pi$ | 43 |
| BaMnSb$_2$ | 22-35 | 0.052-0.058 | 0.21-0.34 | 3.8×10$^{-14}$ | 1280 | 1.06$\pi$ | This work |

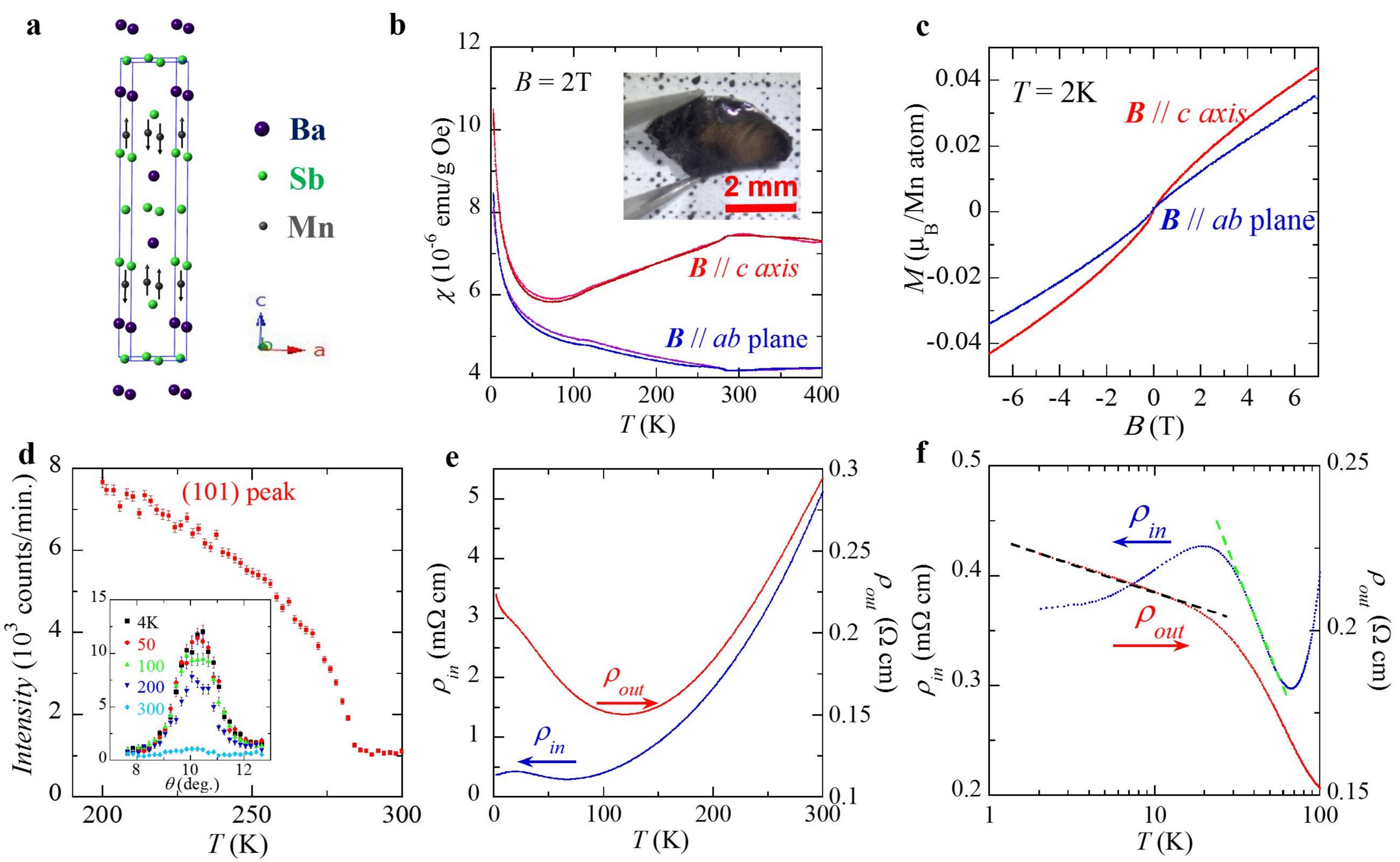

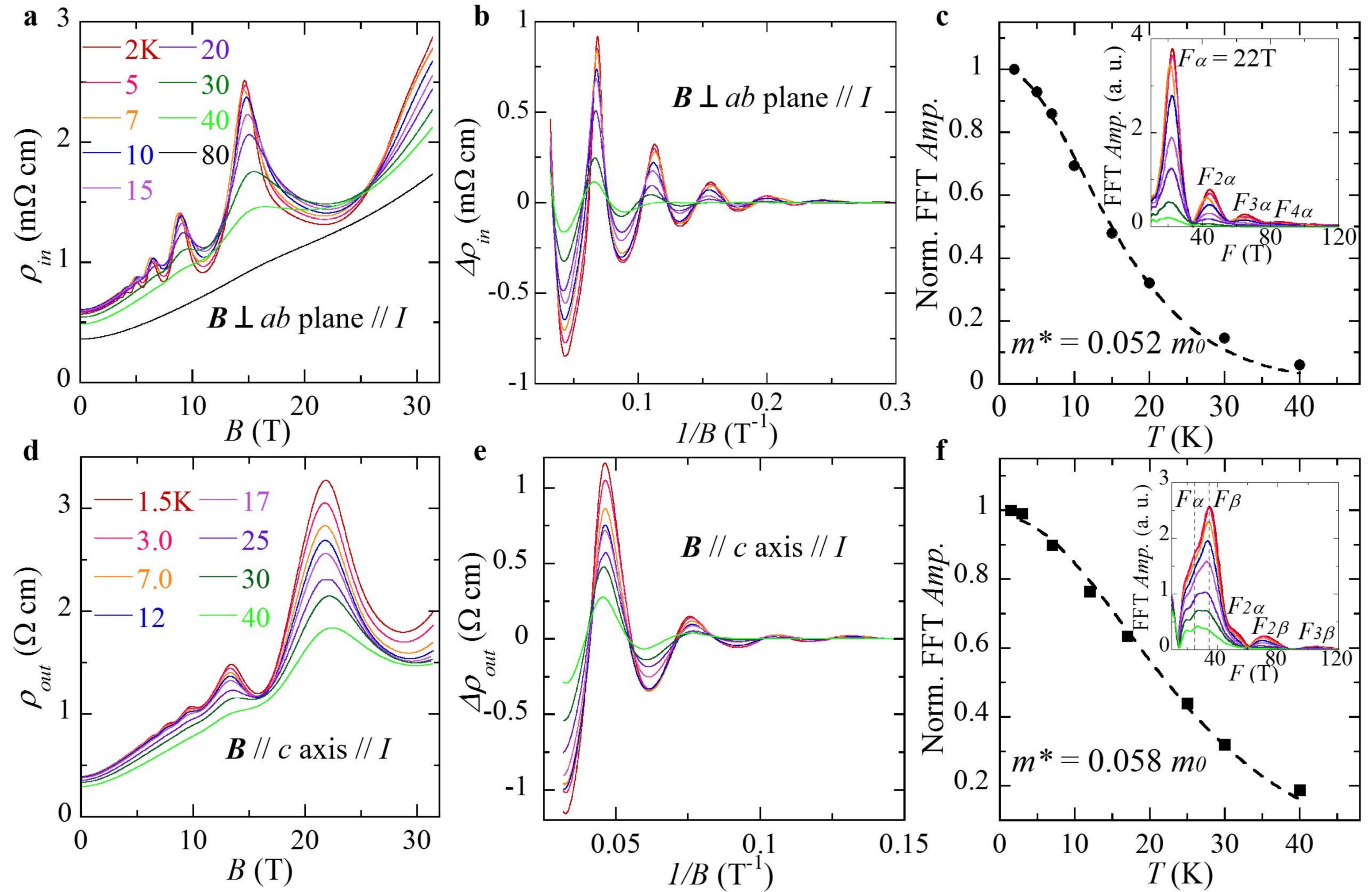

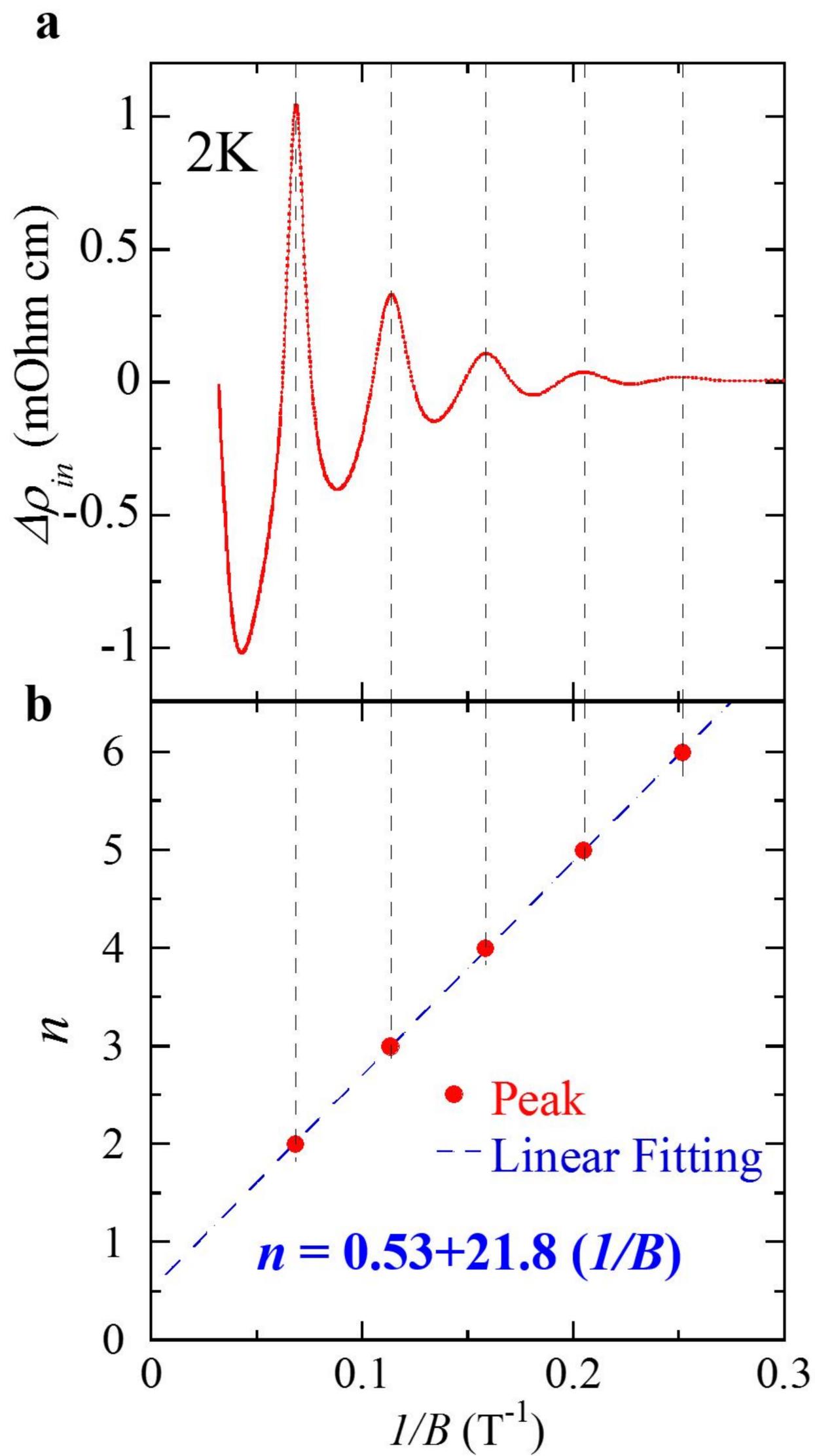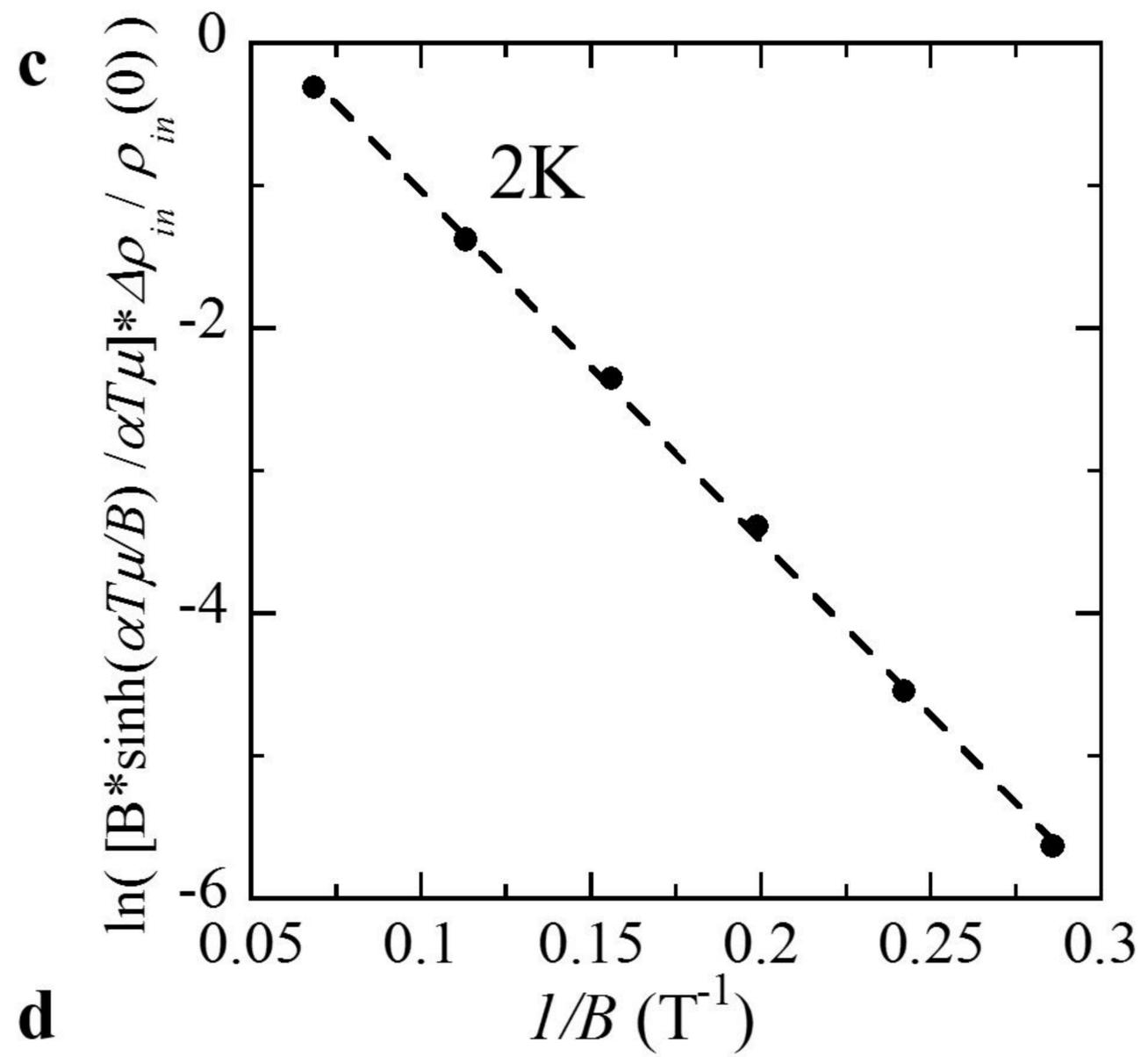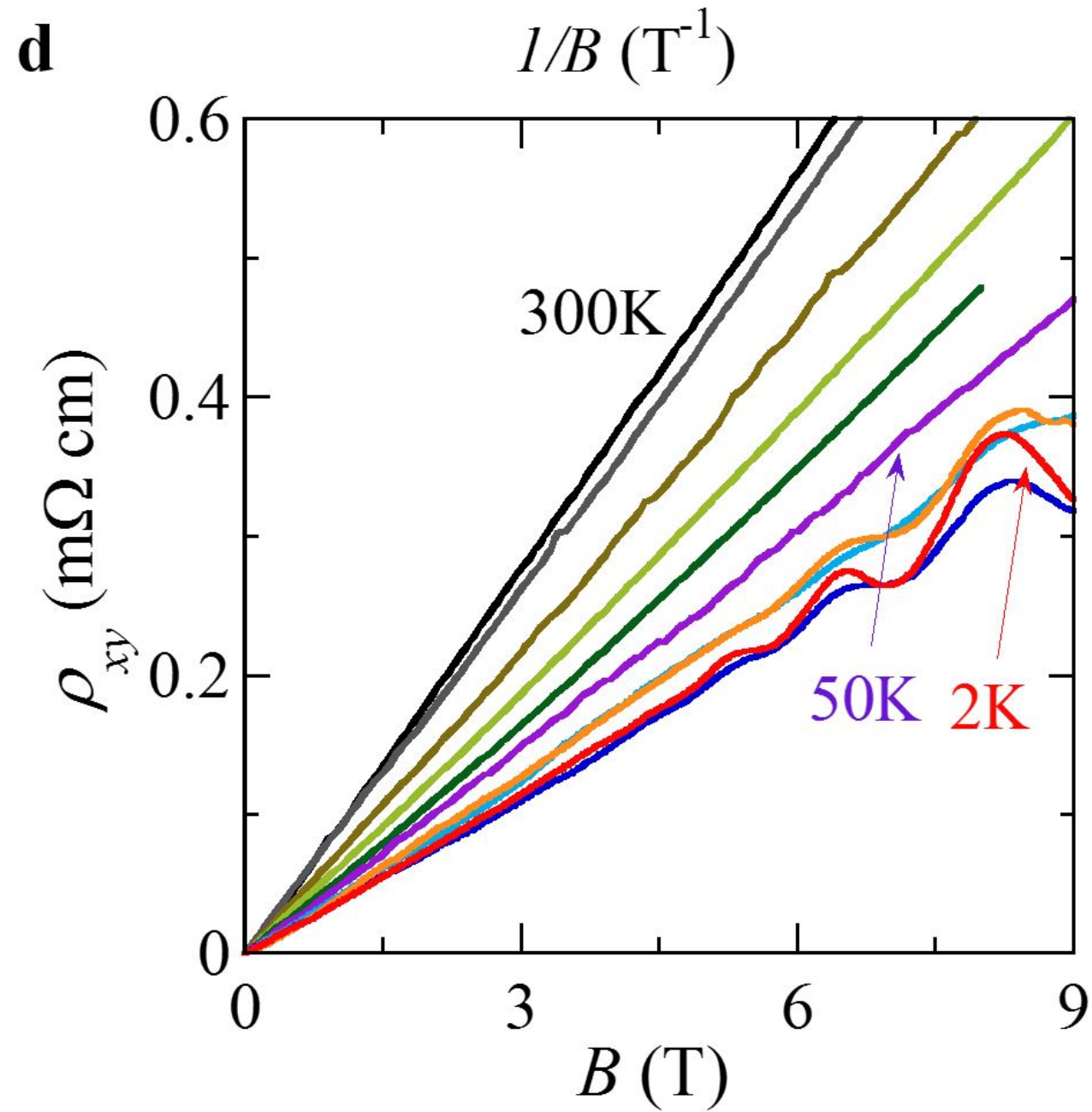

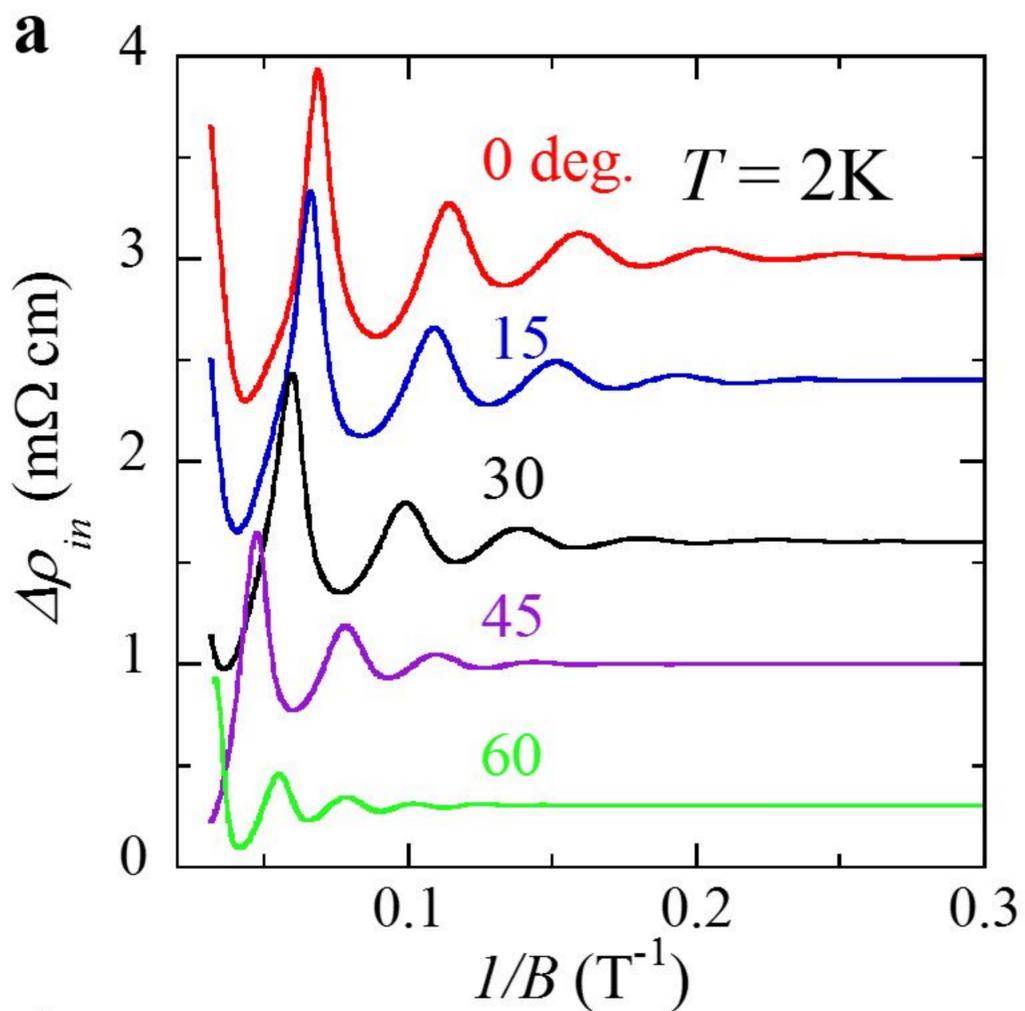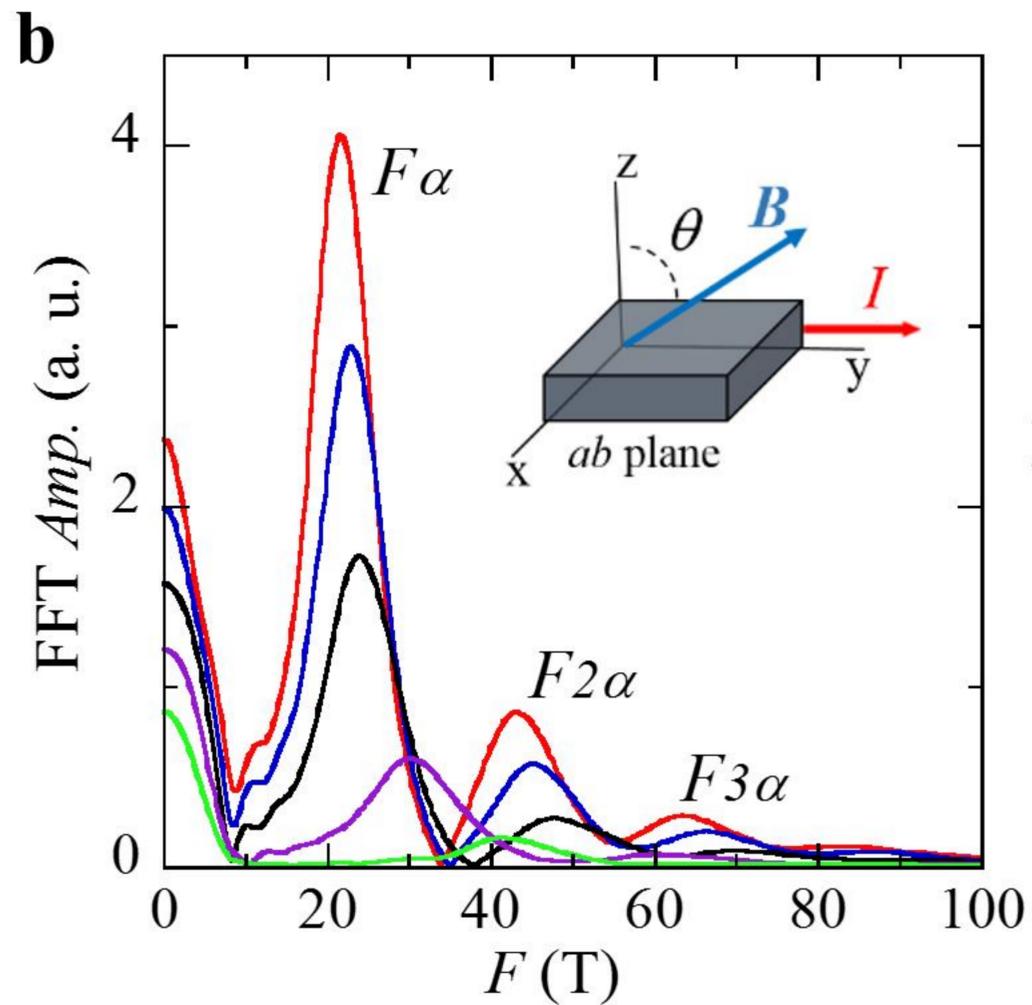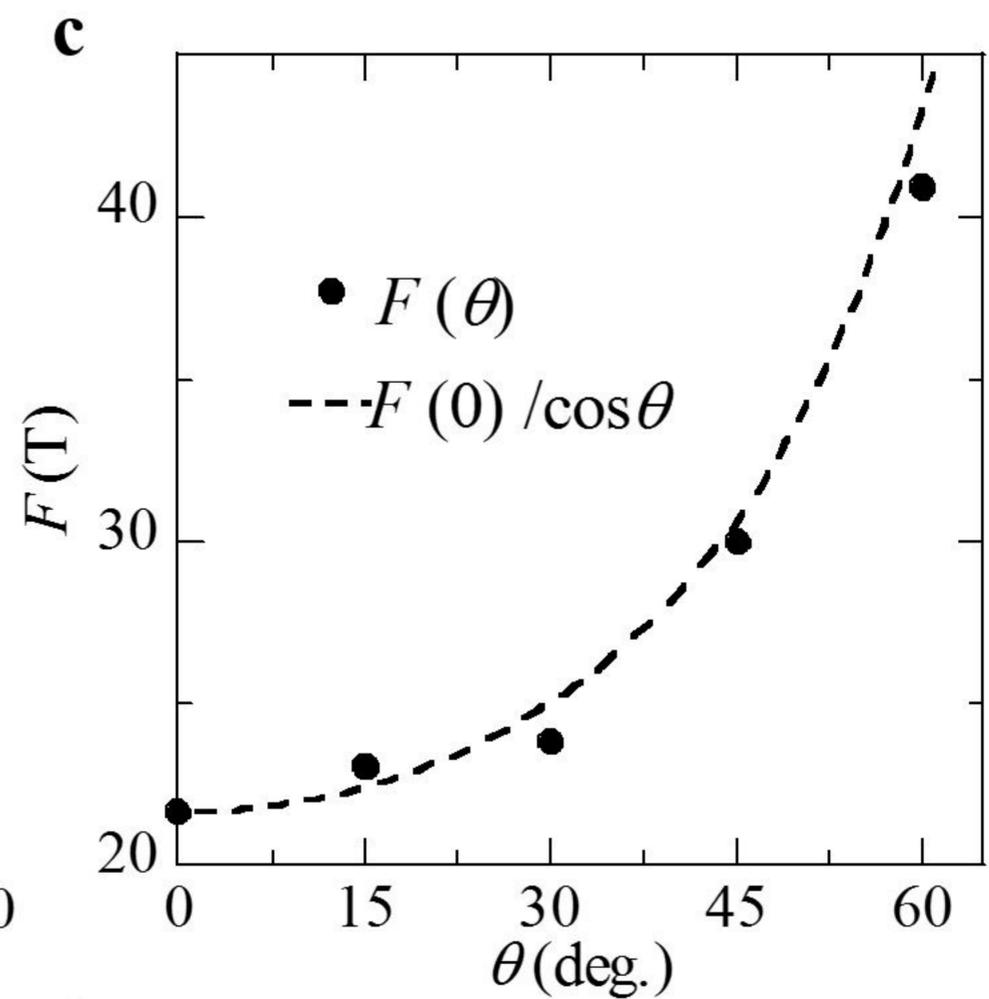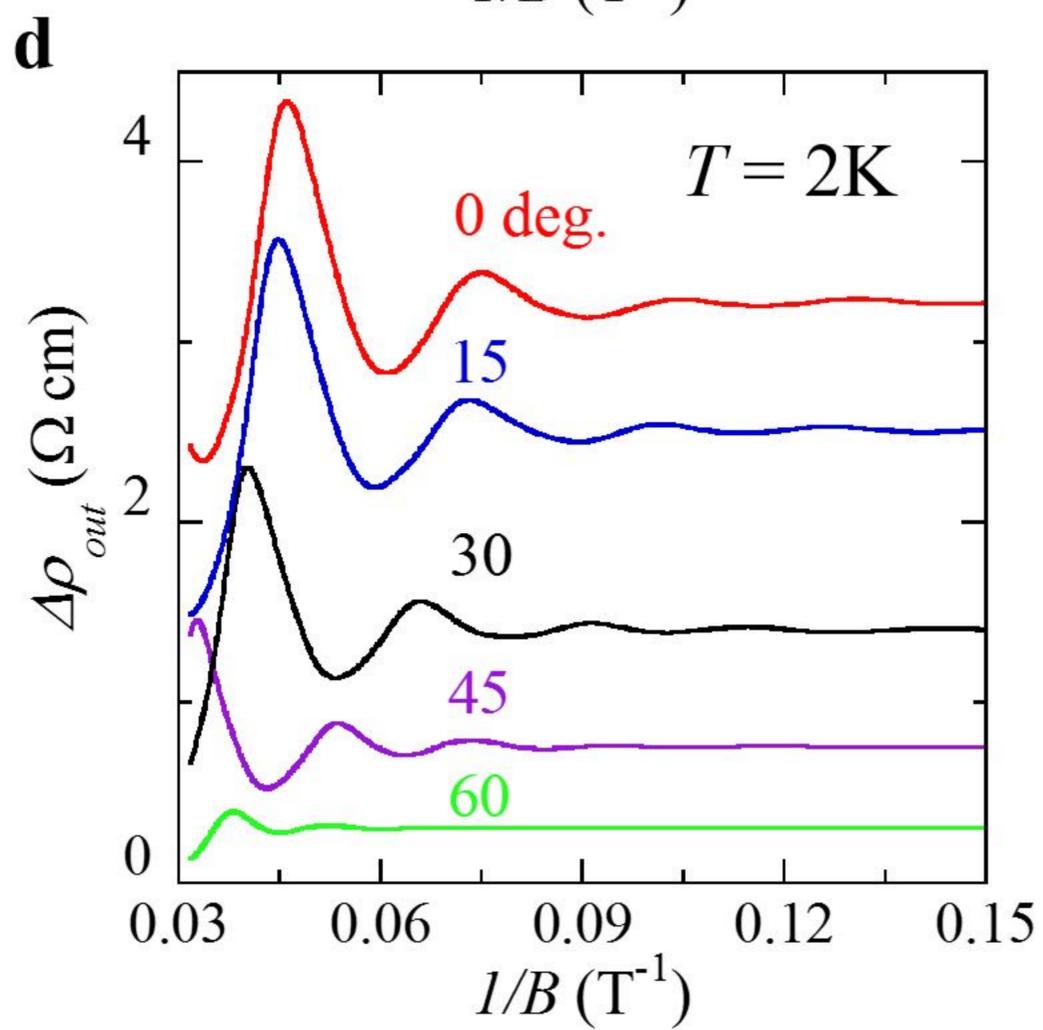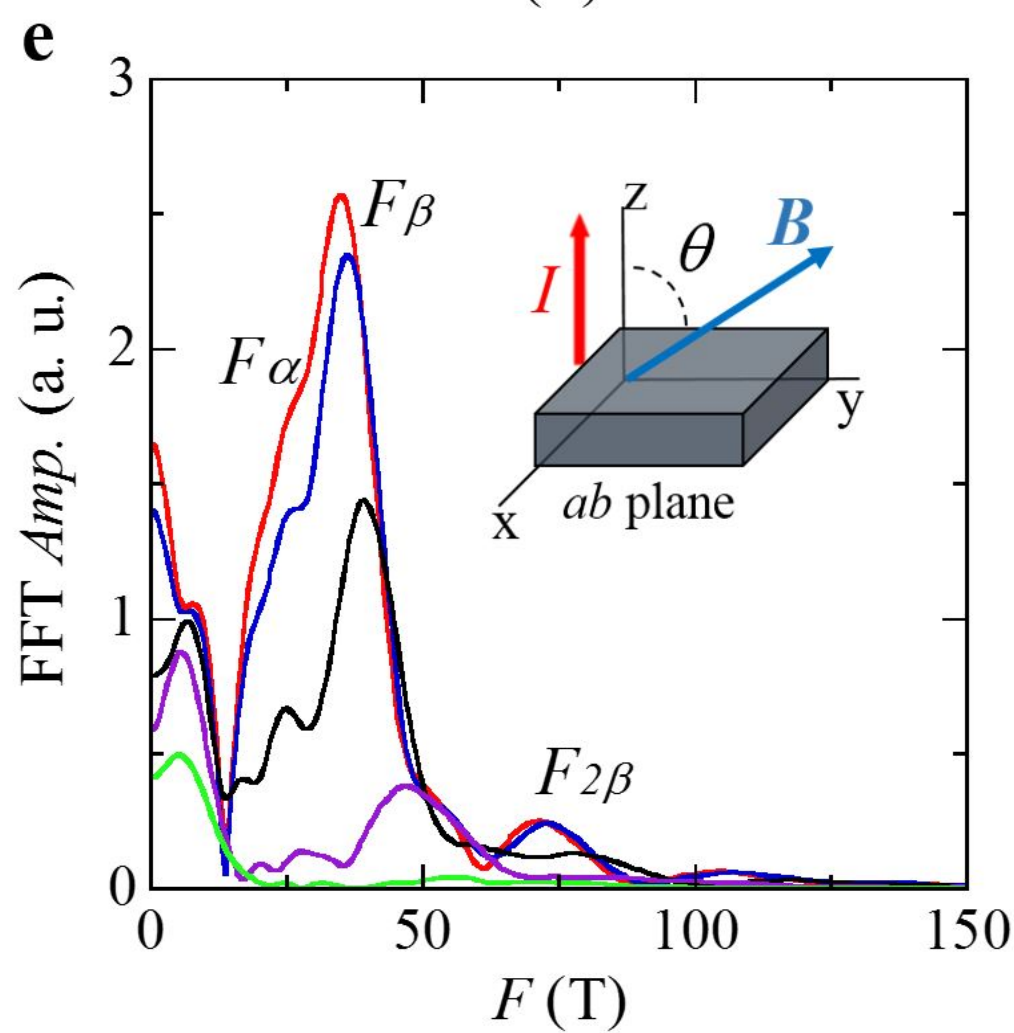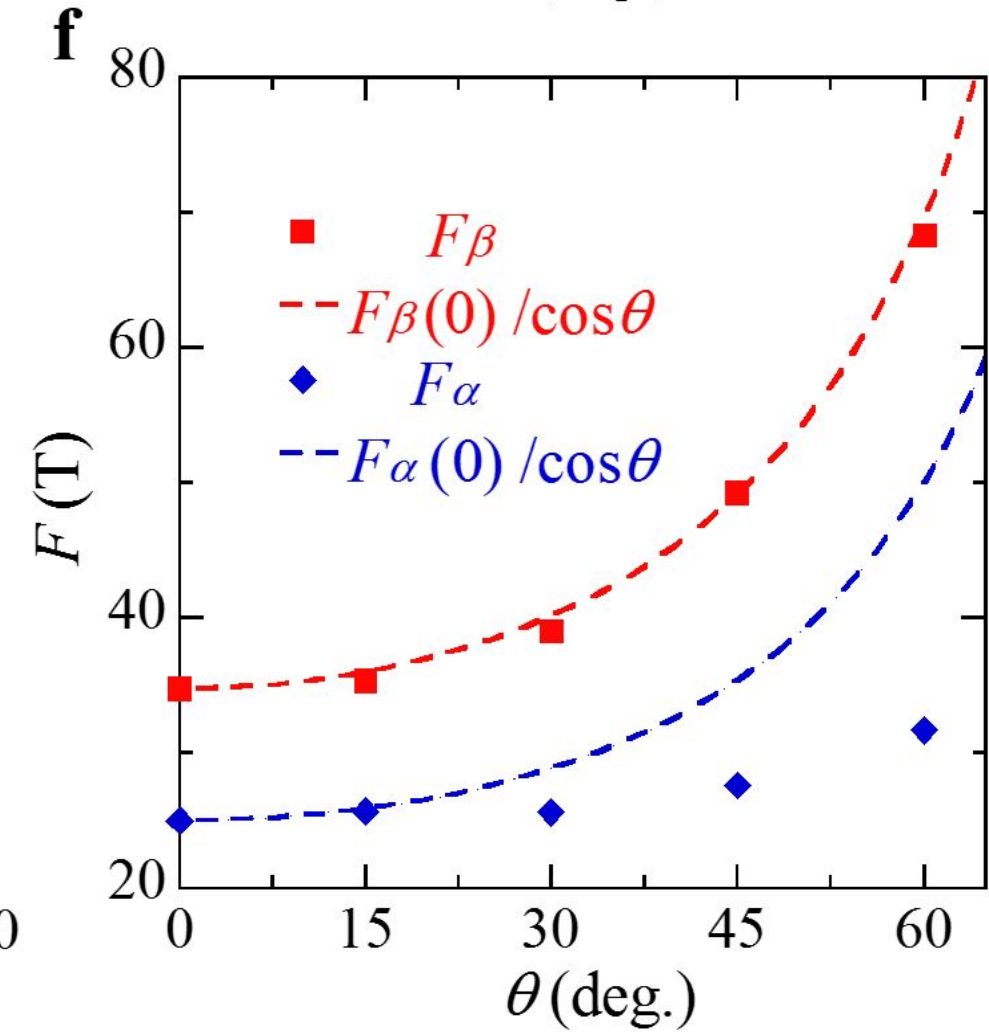